\documentclass{aa}
\usepackage{amsbsy}   
\usepackage{dsfont}
\usepackage{natbib}
\usepackage{url}
\usepackage{graphicx}
\usepackage{relsize}
\usepackage{txfonts}

\bibpunct{(}{)}{;}{a}{}{,} 

\def\newV2{}

\def\hip{{\scshape Hipparcos}}
\def\cpp{C{\relsize{-2}\raise.4ex\hbox{+\kern-.2em+}}}
\def\tyc{\mbox{Tycho\kern.05em-\kern-.1em2}}
\newcommand{\csource}{\texttt}

\def\sind{\sin \delta}       
\def\cosd{\cos \delta}       
\def\sina{\sin \alpha}       
\def\cosa{\cos \alpha}       
\def\mua{\mu_\alpha}         
\def\mud{{\mu_\delta}}       

\def\unit#1{\;{\rm #1}}
\def\masyr{\;{\rm mas\;yr}^{-1}}

\def\la{\mathrel{\mathchoice{\vcenter{\offinterlineskip\halign{\hfil$\displaystyle##$\hfil\cr<\cr\sim\cr}}}{\vcenter{\offinterlineskip\halign{\hfil$\textstyle##$\hfil\cr<\cr\sim\cr}}}{\vcenter{\offinterlineskip\halign{\hfil$\scriptstyle##$\hfil\cr<\cr\sim\cr}}}{\vcenter{\offinterlineskip\halign{\hfil$\scriptscriptstyle##$\hfil\cr<\cr\sim\cr}}}}}

\begin{document}
\title{Formalism and quality of a proper motion link with extragalactic objects for astrometric satellite missions}
\titlerunning{Proper motion link with extragalactic objects for astrometric satellite missions}
\author{M. Metz, M. Geffert}
\authorrunning{M. Metz \& M. Geffert}
\institute{Sternwarte der Universit\"at Bonn, Auf dem H\"ugel 71, 53121 Bonn, Germany}
\date{Received 26 May 2003 / Accepted 25 September 2003}

\abstract{
The accuracy of the link of the proper motion system of astrometric satellite missions like AMEX and GAIA is discussed. Monte-Carlo methods were used to simulate catalogues of positions and proper motions of quasars and galaxies to test the link. The main conclusion is, that future satellite missions  like GAIA may be ``self-calibrated'' by their measurements of QSOs, while additional measurements from radio stars or HST-data are needed to calibrate the less deep reaching astrometric satellite missions of AMEX type.
\offprints{M. Metz, \email{mmetz@astro.uni-bonn.de}}
}
\maketitle

\section{Introduction}\label{sec_intro}
Since 1998 the reference frame for astrometric measurements is represented by the International Celestial Reference Frame (ICRF, \citealt{ma98}), which is the realisation of the International Celestial Reference System (ICRS, \citealt{ari95}), defined by the International Earth Rotation Service (IERS). The ICRF consists of 212 VLBI observed quasars as defining extragalactic radio sources. These objects are assumed to have no proper motion and no overall global rotation -- the quasars represent a quasi inertial system.

In the optical, the ICRS is represented today by the \hip\ catalogue \citep{hipmanu}. To be consistent, the \hip\ catalogue had to be related to the ICRF. This means that it had to be assured that $1^{\mathrm{st}}$, the coordinate axes of the \hip\ system are aligned with those of the ICRF and $2^{\mathrm{nd}}$, that the coordinate axes have no additional time dependent rotation component. Differences would show up as a global offset in positions and in a fictitious proper motion component. 
The transformation between the two systems is in general given by a rotation of the positional and proper motion system. This is the so-called \emph{link algorithm}.

For the \hip\ satellite it was not possible to link the catalogue directly with extragalactic objects, because the only extragalactic object in the \hip\ catalogue is the quasar 3C\,273 (HIP 60936). Consequently intermediate observations were needed \citep{lk97}: VLBI, MERLIN and VLA observations of radio emitting stars, proper motions determined with the HST fine guidance system, and photographic observations with respect to extragalactic objects made for example in Bonn \citep{gef97, tuc97}. The final link solutions were accurate to $0.6\unit{mas}$ for the positions and $0.25 \masyr$ for the proper motion system \citep{lk97}.

%
%
%
%
\newV2{
The present study was initiated for the astrometric satellite mission AMEX\footnote{The missions DIVA \citep{roe97} and FAME \citep{1998AAS...193.1206H} are both cancelled, but the new satellite mission AMEX is a joint American-German merger of them.}, now proposed by the US Naval Observatory, and the ESA GAIA mission \citep{per01}, with a further step in astrometric accuracy, foreseen to be launched no later than 2012.
}

With the expected performance of AMEX some hundreds or even thousands of QSOs will be detectable, for GAIA even about 500\,000 QSOs. If extragalactic objects are detectable it is possible to calibrate the proper motion system without further observations, so this would be a kind of self-calibration.

The aim of this paper is to determine the accuracy of the link of the proper motions of the astrometric satellite missions. We present Monte-Carlo simulations of observations of QSOs with the performance expected for the AMEX satellite and also for the GAIA satellite mission. These simulations allow us to estimate the accuracy of a proper motion link using extragalactic objects.

In Sect.~\ref{sec_linkalgo} the link algorithm will be discussed. Some details about the implementation are described in Sect.~\ref{ssec_implement}. In Sect.~\ref{sec_gal_search} we present a search program for galaxies that could also be used for a proper motion link. In Sect.~\ref{sec_mc_models} we give a brief discussion of the underlying model to simulate the observations and the testing is described in Sect.~\ref{sec_mc_test}. Finally, in Sect.~\ref{sec_mc_sat} we present the results of our simulations.

\section{The link algorithm}\label{sec_linkalgo}
In this section we will give a complete mathematical description of the approximation of a small angle rotation and how it can be used to determine the components of the rotation vectors. 

The link algorithm has been used in several publications like \citet{fr82}, \citet{ari88} or \citet{len97}. In \citet{lk95} a theoretical study of the underlying exact model can be found. However, in most of these papers an approximation algorithm was used and the correction terms that will be described later (Eq. \ref{eqLM21}) were not taken into account, especially for the \hip\ link program. We will show that with improved accuracy of astrometric data it is important to include these correction terms.

The basic idea of the algorithm is that in general the transformation between any two coordinate systems with the same origin is given by a rotation. If the coordinate axes of these systems are nearly coincident a first order approximation of the rotation can be used. It is required that the catalogues representing these two systems have a high degree of internal consistency. This is the case for the ICRF, for the \hip\ catalogue, and it will be for future astrometric catalogues. 

\subsection{The first-order approximation}
Let $\boldsymbol{x}$, $\boldsymbol{y}$ and $\boldsymbol{z}$ be the axes of a cartesian coordinate system: $\boldsymbol{x}$ points towards the origin of right ascension, $\boldsymbol{z}$ towards the celestial north pole and $\boldsymbol{y} = \boldsymbol{z} \times \boldsymbol{x}$. Following \citet{mu83}, the coordinates of a unit vector $\boldsymbol{r}$ towards a celestial object in this coordinate system are given in general by
\begin{equation}
\boldsymbol{r} =
\left(\begin{array}{c}
\cosd \cdot \cosa \\ \cosd \cdot \sina \\ \sind
\end{array}\right).
\end{equation}
The two orthogonal vectors $\boldsymbol{p}$ and $\boldsymbol{q}$ which point towards positive right ascension and declination, respectively, are given by
\begin{equation}
\boldsymbol{p} = \langle \boldsymbol{z} \times \boldsymbol{r} \rangle 
=\left(\begin{array}{c}
-\sina \\ \cosa \\ 0
\end{array}\right)
\label{eqn_p}
\end{equation}
and
\begin{equation}
\boldsymbol{q} = \boldsymbol{r} \times \boldsymbol{p}
=\left(\begin{array}{c}
-\sind \cdot \cosa \\ -\sind \cdot \sina \\ \cosd
\end{array}\right)
\label{eqn_q}
\end{equation}
where the brackets $\langle \, \rangle$ denotate vector normalisation.

Any small rotation about an arbitrary axis $\boldsymbol{e}$ with value $|\varepsilon| \ll 1$ is given by
\begin{equation}
\Delta\boldsymbol{r} = \boldsymbol{\varepsilon} \times \boldsymbol{r}\label{eqn_rot}
\end{equation}
with the rotation vector $\boldsymbol{\varepsilon}=|\varepsilon| \boldsymbol{e}$. As shown in \citet{mu83} the vector $\boldsymbol{\varepsilon}$ can be split into a rotation $\Delta \alpha$ about $\boldsymbol{z}$ and a rotation $-\Delta \delta$ about $\boldsymbol{p}$:
\begin{equation}
\boldsymbol{\varepsilon} = \Delta\alpha \; \boldsymbol{z} - \Delta\delta \; \boldsymbol{p}\\
\end{equation}
$\Delta\alpha$ and $\Delta\delta$ are observable quantities: differences between right ascension and declination of two astrometric catalogues. 
It then follows
\begin{equation}
\Delta\boldsymbol{r} = \Delta\alpha \; \cosd \; \boldsymbol{p} + \Delta\delta \; \boldsymbol{q}
\end{equation}
and with the Eqs. (\ref{eqn_p}) and (\ref{eqn_q}), one easily finds 
\begin{eqnarray}
\Delta\alpha \cosd &=& \Delta \boldsymbol{r} \cdot \boldsymbol{p}\nonumber\\
&=& - \sind \cosa \; \varepsilon_1 - \sind \sina \; \varepsilon_2 + \cosd \; \varepsilon_3 \label{eqn_Da}\end{eqnarray}
and
\begin{equation}
\Delta\delta = \Delta \boldsymbol{r} \cdot \boldsymbol{q} = 
\sina \; \varepsilon_1 - \cosa \; \varepsilon_2 \label{eqn_Dd}
\end{equation}
where $\varepsilon_i$ denotes the component of the vector $\boldsymbol{\varepsilon}$. 
For the proper motion components one finds
\begin{eqnarray}
\Delta \mua &=& - \cos^2 \delta \cosa \; \mud \; \varepsilon_1 + \tan \delta \sina \; \mua \; \varepsilon_1 \nonumber\\
 &&- \cos^2 \delta \sina \; \mud \; \varepsilon_2 - \tan \delta \cosa \; \mua \; \varepsilon_2 \nonumber\\
 && - \tan \delta \cosa \; \omega_1 - \tan \delta \sina \; \omega_2 + \omega_3 \label{eqn_Dma}
\end{eqnarray}
and
\begin{eqnarray}
\Delta\mud &=& \cosa \; \mua \varepsilon_1 + \sina \; \mua \varepsilon_2 \nonumber\\
 && + \sina \; \omega_1 - \cosa \; \omega_2 \label{eqn_Dmd}
\end{eqnarray}
with $\dot{\alpha}=\mua$, $\dot{\delta}=\mud$ and $\dot{\varepsilon_i}=\omega_i$.

Finally the Eqs. (\ref{eqn_Da}) -- (\ref{eqn_Dmd}) can be written in a matrix equation that can be solved for the components of the rotation vector $\boldsymbol{\varepsilon}$ and the spin vector $\boldsymbol{\omega}$
\begin{equation}
\left(\begin{array}{c}
\Delta \alpha \cosd\\
\Delta \delta \\
\Delta \mua \cosd \\
\Delta \mud
\end{array}\right) = 
\left(\begin{array}{cc}
A_{11} & 0 \\
A_{21} & A_{11}
\end{array}\right)
\;
\left(\begin{array}{c}
\varepsilon_1\\
\varepsilon_2\\
\varepsilon_3\\
\omega_1\\
\omega_2\\
\omega_3
\end{array}\right)\label{eqn_link}
\end{equation}
with the sub-matrices
\begin{equation}
A_{11}=
\left(\begin{array}{ccc}
-\sind\cosa & - \sind\sina & \cosd\\
\sina & -\cosa & 0
\end{array}\right)\label{eqLM11}
\end{equation}
and
\begin{equation}
A_{21}= \left(\begin{array}{ccc}
\left(\begin{array}{cc}- \frac{\cosa}{\cosd} \; \mud +\\ \sind \sina \; \mua \end{array}\right) & 
\left(\begin{array}{cc}- \frac{\sina}{\cosd} \; \mud -\\ \sind \cosa \; \mua \end{array}\right) & 0 \\
\cosa \; \mua & \sina \; \mua & 0
\end{array}\right)\label{eqLM21}
\end{equation}
The Equation (\ref{eqLM21}) is important: this term corrects for the orientation difference of the vectors $\boldsymbol{p}$ and $\boldsymbol{q}$ on slightly different places on the celestial sphere. This is important because these are the directions in which the proper motion components are measured. 

For each celestial object one gets the 4 observational properties and the matrix in Eq. (\ref{eqn_link}) which can be concatenated to a single matrix equation. For values $\Delta X$ we use the convention (reference $-$ satellite) $\Delta X = X_{\rm ref}-X_{\rm sat}$ to calculate the matrix we use the reference catalogue.

\subsection{Solving the equation}\label{ssec_implement}
The matrix equation (\ref{eqn_link}) can be solved by a weighted linear least-squares method. We use the singular value decomposition (SVD) because this algorithm is numerically the most appropriate \citep{nr_cpp}. For the SVD, vector, and matrix operations we use the \cpp\ library \csource{newmat10A} by R. Davies\footnote{Source code available for free from \url{http://www.robertnz.net/nm10.htm}}. Details about the software implementation can be found in \citet{mm2003}.

\section{Galaxies for a proper motion link}\label{sec_gal_search}
In principle galaxies are, besides quasars, also good candidates for a proper motion link if they are sufficiently far away and well detectable for the instrument. Ideal candidates would be galaxies that have a star-like core region for which the position can be measured by the satellite instrument.

While -- as mentioned above -- only one extragalactic object can be found in the \hip\ catalogue, the Tycho measurements should contain data for several galaxies. Since the \tyc\ Catalogue \citep{hog_00} has a completeness of 90\% at $V=11.5\unit{mag}$, one would expect some bright galaxies in this catalogue -- however, not yet recognised as galaxies. On the other hand these objects may be important, since their positions from the \tyc\ mission may be used as first epoch material. In this paragraph we have identified these galaxies and discuss the possibility of using them for the proper motion link of future satellite missions.

\subsection{Search program}
To identify galaxies in the \tyc\ Catalogue we have compared the positions of all objects in the \emph{\tyc\ main catalogue} and the \emph{\tyc\ supplement~1} \citep{hog_t2} with catalogues of galaxies and quasars available at the CDS FTP-Server (\url{ftp://cdsarc.u-strasbg.fr/}). The most important were the \emph{Updated Zwicky Catalogue} (UZW), the \emph{Catalogue of principal galaxies} (PGC), the \emph{Third Reference Catalogue of Galaxies} (RC3) and the \emph{Arcsecond positions of UGC galaxies} (AUGC) catalogue.

All objects found were further reviewed: we checked each single object with Digitized Sky Survey (DSS) images, whenever possible in different colors. In almost all cases miss-identifications due to stars in front of galaxies could be eliminated. All other objects were further checked with the NASA/IPAC Extragalactic Database (NED, \url{http://nedwww.ipac.caltech.edu/}) and Simbad (\url{http://simbad.u-strasbg.fr/Simbad/}).

Finally the galaxies were classified in two categories: ``galaxies in the \tyc\ Catalogue'' and ``\emph{uncertain} galaxies in the \tyc\ Catalogue''. Objects were classified as uncertain in the case of: galaxies in dense star fields, a comment in the database that a star may be in front of the galaxy, or a clearly visible positional offset in the DSS images. Galaxies found in the \tyc\ Supplement~1 were not classified as certain/uncertain.

\subsection{Galaxies in the \tyc\ Catalogue}\label{sec_tycgal}
\begin{figure}
\resizebox{\hsize}{!}{\includegraphics{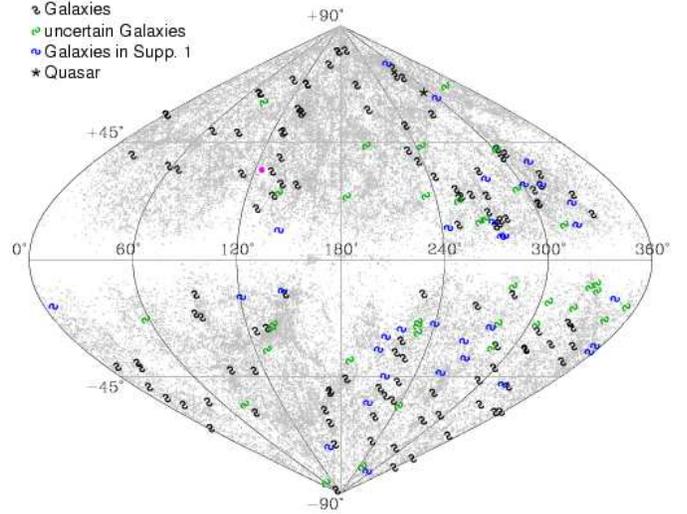}}
\caption{Galaxies in the \tyc\ Catalogue plotted in galactic coordinates. The points in the background are the positions of all PGC galaxies. We have found 116 galaxies, 35 classified as uncertain and 30 galaxies in the \tyc\ Supplement~1. The quasar 3C\,273 was found (again) too.}
\label{fig_tychogal}
\end{figure}
A total number of 181 galaxies were found: 116 galaxies in the \tyc\ Catalogue, 35 uncertain galaxies in the \tyc\ Catalogue and 30 galaxies in the \tyc\ Supplement~1. The tables with all galaxies are available at CDS and the celestial distribution of the galaxies is shown in Fig. \ref{fig_tychogal}. Beside the galaxies, the quasar 3C\,273 that was included in the \hip\ catalogue was found as well.

All galaxies found have a PGC number but not all were found by the position cross correlation with the PGC due to the improper positional data in the PGC (and other galaxy catalogues as well). So there may be more galaxies in the \tyc\ Catalogue which could not be identified by the applied method without more accurate galaxy catalogues.

However the galaxies found are hardly usable as first epoch material for a proper motion determination since they have mean position errors of $\overline{\sigma_{\alpha*}}=117 \pm 22 \unit{mas}$ and $\overline{\sigma_{\delta}}=127 \pm 24 \unit{mas}$. With second epoch material e.g. from GAIA the resulting proper motion error would be of the order of $\sigma_{\mu} \approx 6 \masyr$.  The large errors are mainly due to the fact that the galaxies are close to the limit of the \tyc\ Catalogue with a mean magnitude of $V_{\rm T}=12.5 \unit{mag}$.
\newV2{On the other side one would expect that a few months of observations with GAIA would produce results with a higher accuracy than it could be possible with help of the \tyc\ Catalogue data.}

\subsection{Astrometric Observations}
For three galaxies we made observations at our 1\,m-Cassegrain telescope at the Observatorium Hoher List with the CCD camera HoLiCam \citep{holicam} in October 2002: PGC 7223, 11404, and 65866. For these galaxies we determined the internal position accuracy with a software package developed by \citet{gef97}, by fitting a star profile to the galaxy position. As reference we used the \tyc\ Catalogue. For PGC 7223 and PGC 65866 the positions could be determined equally good as for stars in the fields with mean errors $\la 15\unit{mas}$. Both galaxies have a very bright, star-like core region. For PGC 11404 we found a superimposed faint star but with a position clearly offset from the centre of the galaxy and from the \tyc\ Catalogue position.

The results give a hint that in principle the positions of most of the galaxies found could be determined quite well even if a star profile were fitted, as we did.

\section{Monte-Carlo Simulations: Models and tests}\label{sec_montecarlo}
\subsection{Models for the simulations}\label{sec_mc_models}
To analyse the possibility of linking the proper motion system of an astrometric satellite with QSOs we made Monte-Carlo simulations of observations of QSOs and for comparison also of observations of radio stars. The steps of the simulation were as follows:
\begin{enumerate}
\item[A.] Creation of a (synthetic) reference catalogue.
\item[B.] Creation of a satellite catalogue by applying an exact rotation model.
\item[C.] Applying an error model for the reference catalogue.
\item[D.] Applying an error model for the satellite catalogue.
\item[E.] Using the link algorithm to obtain the rotation and spin vectors.
\item[F.] Comparison with the vectors used in step B.
\end{enumerate}

To create a reference catalogue we could either use a totally synthetic model that creates equally distributed \newV2{apparent brightnesses} and proper motions or make use of existing catalogues to create lists of objects with adopted \newV2{apparent brightness} and proper motion distributions. The first model is useful for testing purposes, the second one for simulations of observations. To apply a rotation to the synthetic reference catalogue we used the exact rotation equations \citep{lk95}.

Errors could be simulated with the following models: no errors; normally distributed errors for testing; 
errors depending on the apparent brightness of a simulated object should be used to simulate the performance of an astrometric satellite (see Sect. \ref{sec_mc_sat}); error distribution of existing catalogues could be used to simulate a realistic error distribution for example of radio observations. The adopted errors were applied to the catalogues by assuming each single error to be normally distributed around zero.\footnote{This does not imply that the total error distribution must be normal distributed!}

We allowed different parameters for the simulations: total number of simulations $N_{\rm sim}$, number of simulated catalogues $N_{\rm cat}$, and number of simulated objects $N_{\rm obj}$. For each simulation we calculated the RMS of the difference between input and derived rotation and spin vector, $\sigma (\varepsilon_i)$ and $\sigma (\omega_i)$, which are a measure for the mean expected errors of each vector component.

\subsection{Testing the algorithm}\label{sec_mc_test}
\begin{figure}
\resizebox{\hsize}{!}{\includegraphics{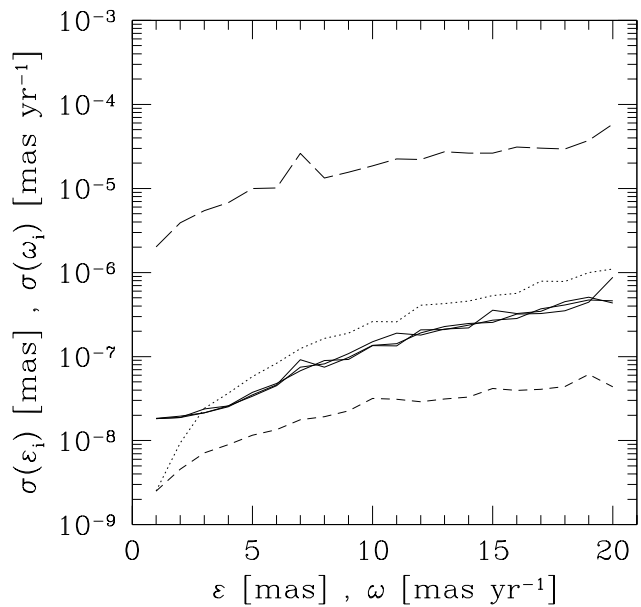}}
\caption{Testing the algorithm performance: the three solid lines show the mean errors of the position link $\sigma(\varepsilon_i)$, the broken lines the mean errors of the proper motion link $\sigma(\omega_i)$, when different parameters were varied: $\varepsilon$ and $\omega$ varied (dotted), only $\omega$ varied (short dashed) and correction terms unused (long dashed lines). The errors of the position link $\sigma(\varepsilon_i)$ are not affected by the variations of $\varepsilon$ or $\omega$.}
\label{fig_sim1}
\end{figure}
We made detailed tests of the simulations and the algorithm. The accuracy of the link algorithm itself could be tested with simulations without errors. We varied the following parameters:
\begin{itemize}
\item number of objects $N_{\rm obj}$
\item absolute value of the rotation and spin vectors $|\varepsilon|$ and $|\omega|$
\item the absolute value of the simulated proper motion of objects $\overline{\mu}$
\end{itemize}
Varying the number of objects, $N_{\rm obj}$, does not lead to a change in accuracy.

The effects of the correction terms of Eq. (\ref{eqLM21}) were tested by varying $|\varepsilon|$ and $|\omega|$ with fixed mean proper motions, which is shown in Fig. \ref{fig_sim1}. The error of the position link $\sigma (\varepsilon_i)$ increases with increasing $|\varepsilon|$. 
The value $\sigma (\omega_i)$ increases similarly if $|\varepsilon|$ and $|\omega|$ increase simultaneously. If in contrast the correction terms are \emph{not} taken into account, the error $\sigma (\omega_i)$ is larger by a factor $\approx 500$. The error $\sigma (\omega_i)$ increases also if only the rotation vector $|\varepsilon|$ is varied whereas the value $\sigma (\varepsilon_i)$ is not affected by any variation of $|\omega|$.

\begin{figure}
\resizebox{\hsize}{!}{\includegraphics{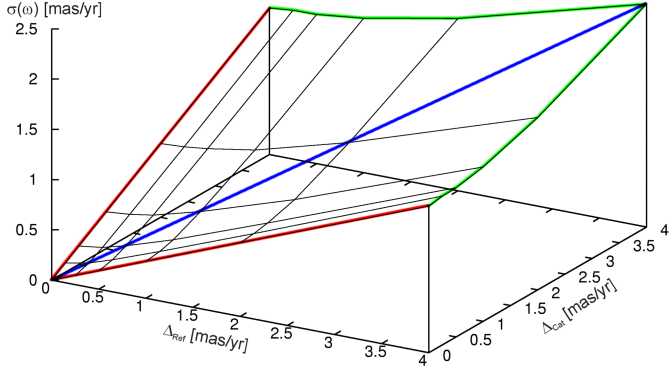}}
\caption{Testing the error behaviour with normally distributed errors for the simulated reference and satellite catalogues. The mean proper motion error $\sigma(\omega_i)$ of the link depends on the size of the mean errors in the reference catalogue $\Delta_{\rm Ref}$ and the satellite catalogue $\Delta_{\rm Cat}$. The surface of error values is, of course, concave.}
\label{fig_3dplot}
\end{figure}

The different behaviour for small values of $|\varepsilon|$ and $|\omega|$ is a numerical effect. Important for the numerical accuracy of the link algorithm are differences between two values, for example $\Delta \alpha$. Whereas for the proper motions one has to store small differences of small values, for positions one has to store small differences of large values. The smallest absolute value $|\Delta \alpha|$ that could be stored in a double precision floating point number for a object with $\alpha>15\fh3$ is $|\Delta \alpha| = 2^{3-52} \approx 4 \cdot 10^{-7} \unit{mas}$.\footnote{For a IEC/IEEE 64 bit floating point number, e.g. a \csource{double} in \cpp\ on i386 architecture.}

Also visible in the plot are little ``bumps''. These appear because the RMS varies due to the Poisson-like distribution while a median would be more appropriate. However, in the plot we show the RMS because this it better comparable with the further results and shows the same basic behaviour as a median.

Further tests showed that the mean errors are independent of the value of the mean proper motion $\overline{\mu}$ if the correction terms are used. If the simulated objects have no proper motion the results do not differ whether the correction terms are taken into account or not.\\

We also made simulations with normally distributed errors for positions and proper motions of both catalogues. If we fix the mean errors of both catalogues and vary the number of link objects $N_{\rm obj}$ the resulting error could, as one would expect, be fitted by a function $\propto 1/\sqrt{N_{\rm obj}}$.

The relation between the resulting link mean error and the mean errors of the catalogue is shown in Fig. \ref{fig_3dplot}. We plotted the error $\sigma(\omega_i)$ depending on the size of the mean errors $\Delta$ of the simulated reference (ref) and satellite (cat) catalogues. The behaviour is as follows: if we fix the mean error of one catalogue, for example $\Delta_{\rm cat}=0\masyr$, and increase the other one, the resulting error for the proper motion link $\sigma(\omega_i)$ increases linearly. If we now fix $\Delta_{\rm ref}=4\masyr$ and increase $\Delta_{\rm cat}$, one sees that the link error is further on dominated by the much larger error $\Delta_{\rm cat}$ until the value of $\Delta_{\rm ref}$ gets in the order of $\Delta_{\rm cat}$. Finally, both contribute equally to the resulting error.

If we assume normally distributed errors for both catalogues a $\chi^2$-Test could be applied, but it fails if we use models for astrometric satellites (see next section), because the errors are no longer normally distributed, which is a requirement for the $\chi^2$-Test \citep{br99}.

\section{Simulations for astrometric satellite missions:\\AMEX like satellites and GAIA}\label{sec_mc_sat}
In \citet{dv_hi98} the relation between the \newV2{apparent brightness} of a star and the parallax accuracy for the DIVA satellite 
 is presented, which then could be transformed in a position and proper motion accuracy. We fitted two exponential functions through the data points \citep{mail_hi02} to get a smooth relation 
\newV2{
and we use this data as estimation for AMEX like satellites (AMEX model). 
}
For GAIA we used data from \citet{per01} and fitted a single exponential function (GAIA model). We then used these models to simulate a proper motion link with different kinds of objects.

\subsection{Simulations with quasars}\label{sec_sim_qua}
To simulate observations of quasars we used the catalogue \emph{Quasars and Active Galactic Nuclei (QAGN)} \citep{ver01} to create the reference catalogue with a luminosity distribution of quasars as realistic as possible. We choose different limiting magnitudes to determine how the accuracy behaves with satellite performance. All proper motions and their errors were set to $0 \masyr$ for the reference catalogue. This means that the resulting accuracy is only depending on the expected errors of the satellite observations.

\begin{figure}
\resizebox{\hsize}{!}{\includegraphics{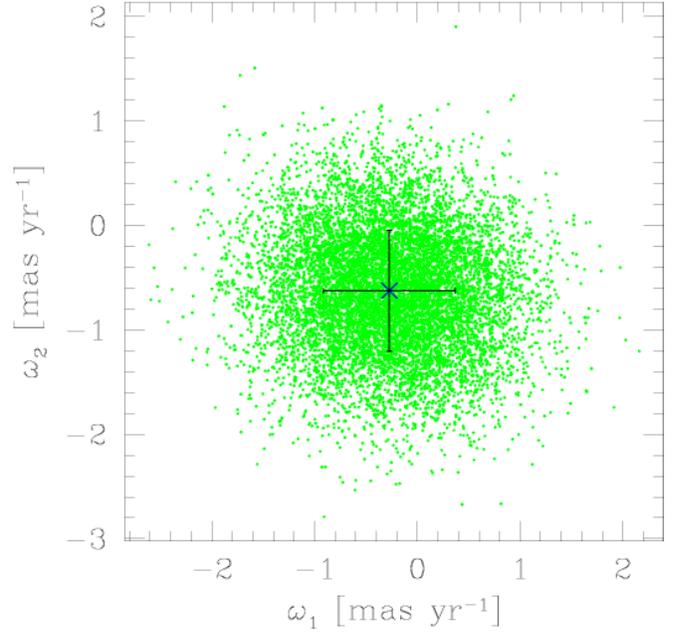}}
\caption{Two components of the rotation vector $\boldsymbol{\omega}$ for a simulation of the proper motion link with quasars with a limiting magnitude of $V_{\rm max}=15\unit{mag}$ and the AMEX model. The components used (cross) are \newV2{reproduced without bias} but the standard deviations are quite large with $\sigma(\omega_1)=0.64\masyr$ and $\sigma(\omega_2)=0.58\masyr$.}
\label{fig_vp}
\end{figure}
The parameters for the AMEX like satellite simulation were a limiting magnitude of $V_{\rm max}=15\unit{mag}$ ($N_{\rm obj}=88$ QSOs) and we made $N_{\rm sim}=10\,000$ simulations. In Fig. \ref{fig_vp} the results of one of those simulations are shown. It is clearly visible that the values used to create the satellite catalogue (marked by the cross) are found on average but the results spread over a large range.

The results of all simulations performed are given in Table \ref{tab_sim_result}. The expected errors of the proper motion link for GAIA are, at the same limiting magnitude and the same number of objects, a factor of $\approx 400$ -- 800 smaller than for the AMEX model, based on the expected GAIA performance \citep{per01}.

The mean error of the proper motion link for the AMEX model is $\sigma(\omega_i)=0.61\masyr$ at a limiting magnitude of $V_{\rm max}=15\unit{mag}$. If we include more QSOs by using fainter objects the mean error decreases only slightly. This is due to the fact that simultaneously the error of the proper motion determination increases for fainter objects. So the profit of using more objects is essentially cancelled out by the worsened performance. For GAIA the error decreases steadily.

To study the influence of the individual parameters we have also varied the number of QSOs and the length of the  mission, both increased by a factor of 1.5. The results were as expected: if we increase the number of objects by a factor of 1.5 the error of the proper motion link decreases by a factor of $\sqrt{1.5}$, \newV2{whereas a 1.5 times longer mission leads to a link error that is by a factor $\sqrt{1.5^3}$ smaller \citep{dv_hi98}}.

\newV2{
In \citet{kov03} the effect of the curvature of the Solar system barycentric motion around the centre of our galaxy is described. This appears as an apparent proper motion of up to $4\unit{\mu as\;yr^{-1}}$. Our simulations show that this affects the proper motion link by an error that is a factor of $\approx$ 10\% larger for all limiting magnitudes for the GAIA model if the data is not corrected for the aberration effect (which can be done before the link is applied). The apparent proper motion has no significant influence on the AMEX model with quasars.
}

It has to be taken into account that the QAGN is not complete for faint objects. If we extrapolate the number of quasars brighter than $V_{\rm max} = 16.5 \unit{mag}$ in the \emph{SDSS Quasar Catalog I} \citep{sn02} we find that at this limiting magnitude about $57\% \pm 16\%$ of the quasars are in the QAGN whereas only $5.4\% \pm 0.1\%$ brighter than $V_{\rm max} = 19 \unit{mag}$ are in the QAGN. So for GAIA we underestimate the number of quasars in our simulations by a large factor.

\begin{table}
\caption{Expected mean errors $\sigma(\omega_i)$ of the proper motion link when QSOs were used for the simulations with the expected AMEX and GAIA performances. Proper motions and their errors were set to $0\masyr$ for the reference catalogues.}
\centering
\begin{tabular}{cccc}
\hline
\hline
&& AMEX model & GAIA model\\
$V_{\rm max}\unit{[mag]}$ & $N_{\rm obj}$ & $\sigma(\omega_i)\unit{\left[\masyr\right]}$ & $\sigma(\omega_i)\unit{\left[\masyr\right]}$\\
\hline
14   &    28 & 0.65 & 0.0017 \\
14.5 &    49 & 0.63 & 0.0015 \\
15   &    88 & 0.61 & 0.0012 \\
15.5 &   163 & 0.60 & 0.0011 \\
16   &   315 & 0.58 & 0.00094 \\
16.5 &   624 & 0.58 & 0.00083 \\
17   &  1156 & 0.57 & 0.00074 \\
17.5 &  2056 &      & 0.00068 \\
18   &  3453 &      & 0.00063 \\
18.5 &  5542 &      & 0.00060 \\
19   &  8095 &      & 0.00058 \\
19.5 & 11442 &      & 0.00057 \\
20   & 15533 &      & 0.00057 \\
\hline
Galaxies
     &   116 & 0.16 & \\
\hline
\end{tabular}
\label{tab_sim_result}
\end{table}

\subsection{Simulations with galaxies}\label{sec_sim_gal}
As described in Sect. \ref{sec_gal_search} we identified 116 galaxies in the \tyc\ Catalogue. We used the position and magnitude data from the \tyc\ Catalogue to simulate a proper motion link for AMEX like satellites with galaxies. All proper motions and their errors were set to $0 \masyr$ for the reference catalogue, for the satellite errors we used the AMEX model. The resulting mean errors of the proper motion link were $\sigma(\omega_i)=0.16\masyr$.

This result is a crude estimation. Firstly we made the assumption that the astrometric errors only depend on the magnitude but they may be affected by the source not being point-like. Secondly we used \tyc\ magnitudes as estimation of the brightness of the core region. On the other side it is likely that more galaxies may be detected (see Sect. \ref{sec_tycgal}).

\subsection{Simulations with radio stars}\label{sec_sim_rad}
\begin{table}
\caption{Expected mean errors $\sigma(\omega_i)$ of the proper motion link when radio-stars were used for the link with the AMEX model. We used the data of the 11 radio-stars observed with the VLBI and used for the \hip\ link. In column 3 the errors of the VLBI observations are smaller by a factor of 2.}
\centering
\begin{tabular}{ccc}
\hline
\hline
&& 1/2 radio errors\\
$N_{\rm obj}$ & $\sigma(\omega_i)\unit{\left[\masyr\right]}$ & $\sigma(\omega_i)\unit{\left[\masyr\right]}$\\
\hline
10  & 0.076 & 0.047\\
20  & 0.048 & 0.031\\
30  & 0.038 & 0.025\\
40  & 0.033 & 0.022\\
50  & 0.029 & 0.019\\
60  & 0.026 & 0.017\\
70  & 0.024 & 0.016\\
80  & 0.022 & 0.015\\
90  & 0.021 & 0.014\\
100 & 0.020 & 0.013\\
\hline
\end{tabular}
\label{tab_sim_radio}
\end{table}
To compare the results of the simulations with quasars and galaxies with other methods we made simulations with radio-emitting stars \citep{les98} which were used for the \hip\ link program. Radio stars are suitable link objects, because their  proper motions may be determined with high accuracy within short time scales. For the construction of the synthetic reference catalogue the astrometric data from \citet{les98} were used, magnitudes were taken from the \hip\ catalogue. We varied the number of objects $N_{\rm obj}$ and in a second run the VLBI errors were reduced by a factor of 2.

The results are given in Table \ref{tab_sim_radio}. The errors are mainly caused by the VLBI errors. The accuracy of the proper motion link could be increased by using more link objects which could be described by a function $\propto 1/\sqrt{N}$ as shown in Sect. \ref{sec_mc_test} for normal distributed errors.

\section{Conclusions and outlook}
The main goal of the proper motion link of future astrometric satellite missions is to establish a proper motion system with an accuracy that is better than the accuracy of the single proper motions of one star. Our simulations show that for astrometric missions like GAIA this may be achieved by the use of QSOs alone. For astrometric satellite missions with AMEX like performance the use of quasars alone is not sufficient and galaxies have to be included. However, an accurate proper motion link requires external data from VLBI observations of radio-emitting stars or HST observations of stars with respect to extragalactic objects.

\begin{acknowledgements}
We thank the referee Jean Kovalevsky for his useful comments as well as Ulrich Bastian and Klaas S. de~Boer for useful discussions.
\end{acknowledgements}

\bibliographystyle{aa}
\bibliography{zitate}
\end{document}